\documentclass[11pt,twoside]{article}
\usepackage{asp2010}

\resetcounters

\begin{document}

\title{Low--Mass Stars in the Sloan Digital Sky Survey: Galactic
  Structure, Kinematics, and the Luminosity Function}
\author{John J. Bochanski$^{1,2}$
\affil{$^1$Astronomy \& Astrophysics Dept., Pennsylvania State
  University, 525 Davey Laboratory, University Park, PA, 16802, USA,
  email: jjb29@psu.edu}
\affil{$^2$Kavli Institute for Astrophysics and Space Research,
  Massachusetts Institute of Technology, Building 37, 77 Massachusetts
  Avenue, Cambridge, MA 02139, USA}}

\begin{abstract}
Modern sky surveys, such as the Sloan Digital Sky Survey and
the Two--Micron All Sky Survey, have revolutionized the study of
low--mass stars. With millions of photometric and spectroscopic
observations, intrinsic stellar properties can be studied with
unprecedented statistical significance. Low--mass stars dominate the
local Milky Way and are ideal tracers of the Galactic potential and the
thin and thick disks. Recent efforts, driven by SDSS observations,
have sought to place the local
low-mass stellar population in a broader Galactic context.

I highlight a recent measurement of the luminosity and mass functions
of M dwarfs, using a new technique optimized for large surveys.
Starting with SDSS photometry, the
field luminosity function and local Galactic structure are
measured simultaneously. The sample size used to estimate the LF is
nearly three orders of magnitude larger than any previous study,
offering a definitive measurement of this quantity. The observed LF is
transformed into a mass function and compared to previous studies.

Ongoing investigations employing M dwarfs as tracers of
Galactic kinematics are also discussed. SDSS spectroscopy has produced databases
containing tens of thousands of low--mass stars, forming a
powerful probe of the kinematic structure of the Milky Way. SDSS
spectroscopic studies are complemented by large proper motion surveys,
which have uncovered thousands of common proper motion binaries
containing low--mass stars. Additionally, the SDSS spectroscopic data
explore the intrinsic properties of M dwarfs, including metallicity
and magnetic activity.

The highlighted projects demonstrate the advantages and problems with using
large data sets and will pave the way for studies with next--generation
surveys, such as PanSTARRS and LSST.
\end{abstract}

\section{Introduction}
Low--mass dwarfs ($0.08~M_{\odot} < M < 0.8~M_{\odot}$) are the
dominant stellar component of the Milky Way, composing
$\sim 70\%$ of all stars
\citep{1997AJ....113.2246R,2010AJ....139.2679B} and nearly half the
stellar mass of the Galaxy.  However, despite their abundance,  M
dwarfs were studied in relatively small samples, due to their dim
intrinsic brightnesses ($L \lesssim 0.05~L_{\odot}$).  However, the situation
has been radically altered in the last decade, as deep surveys covering
large areas of the sky were carried out.  These projects, such as the
Sloan Digital Sky Survey \citep[SDSS,][]{2000AJ....120.1579Y} and the
Two--Micron All Sky Survey \citep[2MASS,][]{2006AJ....131.1163S}, can
trace their roots back to photographic surveys, epitomized by the
National Geographic Society - Palomar Observatory Sky Survey
\citep[POSS-I,][]{1963bad..book..481M} and its successor, POSS-II
\citep{1991PASP..103..661R}.  However, the photometric precision achieved by
modern surveys distinguishes them from their photographic predecessors.  For
example, SDSS has imaged 1/4 of the sky to $r \sim 22$ and 2MASS
imaged  the entire sky  to $J \sim 16.5$, with typical uncertainties
of a few percent.  The resulting databases
contain accurate  multi-band photometry of tens of millions of
low--mass stars, enabling exciting new science.  While there is a rich
heritage of historical investigations, this article will focus on results derived from SDSS data.

A multitude of studies have focused on the intrinsic properties of
low--mass stars using SDSS observations.   The field luminosity function (LF)
and corresponding mass function (MF) was measured using over 15
million stars \citep{2010AJ....139.2679B}.  The mass--radius relation of M
dwarfs has been studied with eclipsing binary systems \citep{2008ApJ...684..635B,
  2008MNRAS.386..416B}.  Average photometric colors and spectroscopic features have also
been quantified \citep{2002AJ....123.3409H,
  2007AJ....133..531B, 2007AJ....134.2430D,2008AJ....135..785W,
  west10}.  Multiple studies have attempted to estimate the
absolute magnitude of low--mass stars in SDSS
\citep{2002AJ....123.3409H, 2008ApJ...673..864J, 2008ApJ...689.1244S,
  bochanskithesis, 2009MNRAS.396.1589B, bochanski10}.  However, due to the
lack of precise trigonometric parallaxes for many of the M dwarfs in
SDSS, absolute magnitudes and distances are derived by secondary means.
Chromospheric activity, driven by magnetic dynamos within the stars,
has been observationally traced with H$\alpha$ emission measured in
SDSS spectroscopy \citep{2004AJ....128..426W, 2008AJ....135..785W,
  west10,2010ApJ...722.1352K}.  Flare rates have
been measured for thousands of stars \citep{2009AJ....138..633K,
  2010AJ....140.1402H} and will be
incorporated into predicting the flaring population observed by
next--generation surveys \citep{hilton10}.  

While interesting objects in their own right, M dwarfs
are also powerful tools for studying the Milky Way.   Their ubiquity,
combined with lifetimes much greater than a Hubble time
\citep{1997ApJ...482..420L}, make them ideal tracers of Galactic structure and
kinematics.  Using photometry of over 15 million M dwarfs,
\cite{2010AJ....139.2679B} measured the scale heights and lengths of
the thin and thick disks.  A complementary study by
\cite{2008ApJ...673..864J} used higher mass stars to measure the
stellar density profiles.  The local gravitational potential has been
probed using the kinematics of $\sim$ 7000 M dwarfs along one line of
sight with proper motions, SDSS spectroscopy and photometry
\citep{2007AJ....134.2418B}.  This study has been expanded to the
entire SDSS footprint, using a database of $\sim$ 25,000 M dwarfs with
velocities and distances \citep{pineda}.  Furthermore, \cite{2009AJ....137.4149F}
employed proper motions and distances of $\sim$ 2 million M dwarfs to estimate
velocity distributions.

Below, I highlight several studies that have used SDSS
observations to study both the intrinsic properties of low--mass stars
\emph{and} use them to study the Milky Way.  In \S \ref{sec:observations},
the technical details of SDSS photometric and
spectroscopic observations are briefly described.  In \S \ref{sec:lf},
a new investigation measuring the M dwarf field LF and MF and Galactic
structure parameters is detailed.  Kinematics within the Milky Way are
explored in \S \ref{sec:kinematics}.  The importance of placing large,
deep surveys of M dwarfs in a Galactic context is discussed in \S
\ref{sec:case_study}.  Concluding remarks and avenues for future
investigations are discussed in \S \ref{sec:conclusions}. 

\section{Observations}\label{sec:observations}
The Sloan Digital Sky Survey \citep[SDSS,][]{2000AJ....120.1579Y}  was
a photometric and spectroscopic survey covering $\sim$
10,000 sq.\ deg.\ centered on the Northern Galactic Cap conducted with
a 2.5m telescope \citep{2006AJ....131.2332G} at Apache Point
Observatory (APO).  Photometry was acquired in
five filters
\citep[$ugriz$,][]{1996AJ....111.1748F,2007AJ....134..973I} down to a
limiting magnitude $r \sim 22$
\citep{2002AJ....123..485S,2008ApJS..175..297A}.   Calibration to a
standard star network \citep{2002AJ....123.2121S} was obtained through
concurrent observations
with the ``Photometric Telescope'' \citep[PT;
][]{2001AJ....122.2129H,2006AN....327..821T}. Absolute astrometric
accuracy was estimated to better than 0.1\arcsec~\citep{2003AJ....125.1559P}, and
has been used to measure proper motions with a precision of 3 mas
yr$^{-1}$ \citep{2004AJ....127.3034M}.  When the Northern Galactic Cap
was not visible at APO, a 300 sq.\ deg.\ stripe (Stripe 82) centered on zero
declination was scanned repeatedly.  Observations of Stripe 82
were used to empirically quantify photometric precision
\citep{2007AJ....134..973I}.  Over 357 million unique photometric
objects have been identified in the latest public data release
\citep[DR7,][]{2009ApJS..182..543A}.  The photometric precision of
SDSS is unrivaled for a  survey of its size, with typical errors
$\lesssim$ 0.02 mag \citep{2004AN....325..583I,2007AJ....134..973I}. 

When the conditions at APO were not photometric, SDSS operated in a
spectroscopic mode.  Twin fiber--fed spectrographs obtained
medium--resolution ($R = 1,800$) optical spectra, covering 3800-9200
\AA.  Objects with SDSS photometry were targeted for spectroscopic
followup by complex algorithms.  The primary targets were galaxies
\citep{2002AJ....124.1810S} and quasars \citep{2002AJ....123.2945R}.
However, over $\sim$ 70,000 M dwarfs have been identified within the
SDSS spectroscopic database \citep{west10} from targeted and
serendipitous (i.e., failed quasar) observations.   

\section{Luminosity Function and Galactic Structure}\label{sec:lf}
In the following section, the method and results of
\cite{2010AJ....139.2679B} are detailed.  The
\cite{2010AJ....139.2679B} study produced two distinct yet intertwined
results:  it
combined the investigation of an intrinsic stellar property (the MF and LF) with
a measurement of Galactic structure and self--consistently described the local
low--mass stellar population.  The wide, deep, precise photometric
coverage of SDSS enabled this type of analysis.

Previous investigations of the LF and MF have relied on one of two
techniques:  1) nearby, volume--limited studies of trigonometric parallax
stars \citep[e.g.][]{1997AJ....113.2246R}; or 2) pencil--beam surveys of distant stars
over a small solid angle \citep[e.g.][]{2001ApJ...555..393Z}.  In both cases, sample
sizes were limited to a few thousand stars, prohibiting detailed
statistical measurements.    
Using a sample drawn from SDSS, 2MASS and Guide Star Catalog
photometry and supplemented with SDSS spectroscopy, \cite{covey08}
performed the largest field low--mass LF and MF investigation prior to
the \cite{2010AJ....139.2679B} study. \cite{covey08} measured the
LF with a sample covering 30 sq.\ deg.\ and
containing $\sim 30,000$ low--mass stars.  They also
constrained their contamination rate to $\lesssim$ a few percent after obtaining spectra of every red
point source in a 1 sq.\ deg.\ calibration region.


\subsection{Sample Selection \& Method}
Please refer to \cite{2010AJ....139.2679B} for a detailed
description of sample selection.  M dwarfs were selected from SDSS Data Release 6
\citep[DR6;][]{2008ApJS..175..297A} footprint using $r-i$ and $i-z$
photometric colors. Noisy photometric
observations were removed using flag cuts, yielding a final database
containing over 15 million low--mass stars.

The \cite{2010AJ....139.2679B} photometric sample comprised a data set three orders of
magnitude larger (in number) than any previous LF study.  Furthermore,
it covered over 8,400 sq$.$ deg$.$, nearly 300 times larger than the
sample analyzed by \cite{covey08}.  The large sky coverage of SDSS
presented the main challenge in measuring the LF.  Most 
previous studies either assumed a uniform density distribution (for
nearby stars) or calculated a Galactic density profile, $\rho(r)$
along one line of sight.  With millions of stars spread over nearly
1/4 of the sky, numerically integrating Galactic density profiles for
each star was computationally prohibitive.   

\begin{figure}[!ht]
 \plottwo{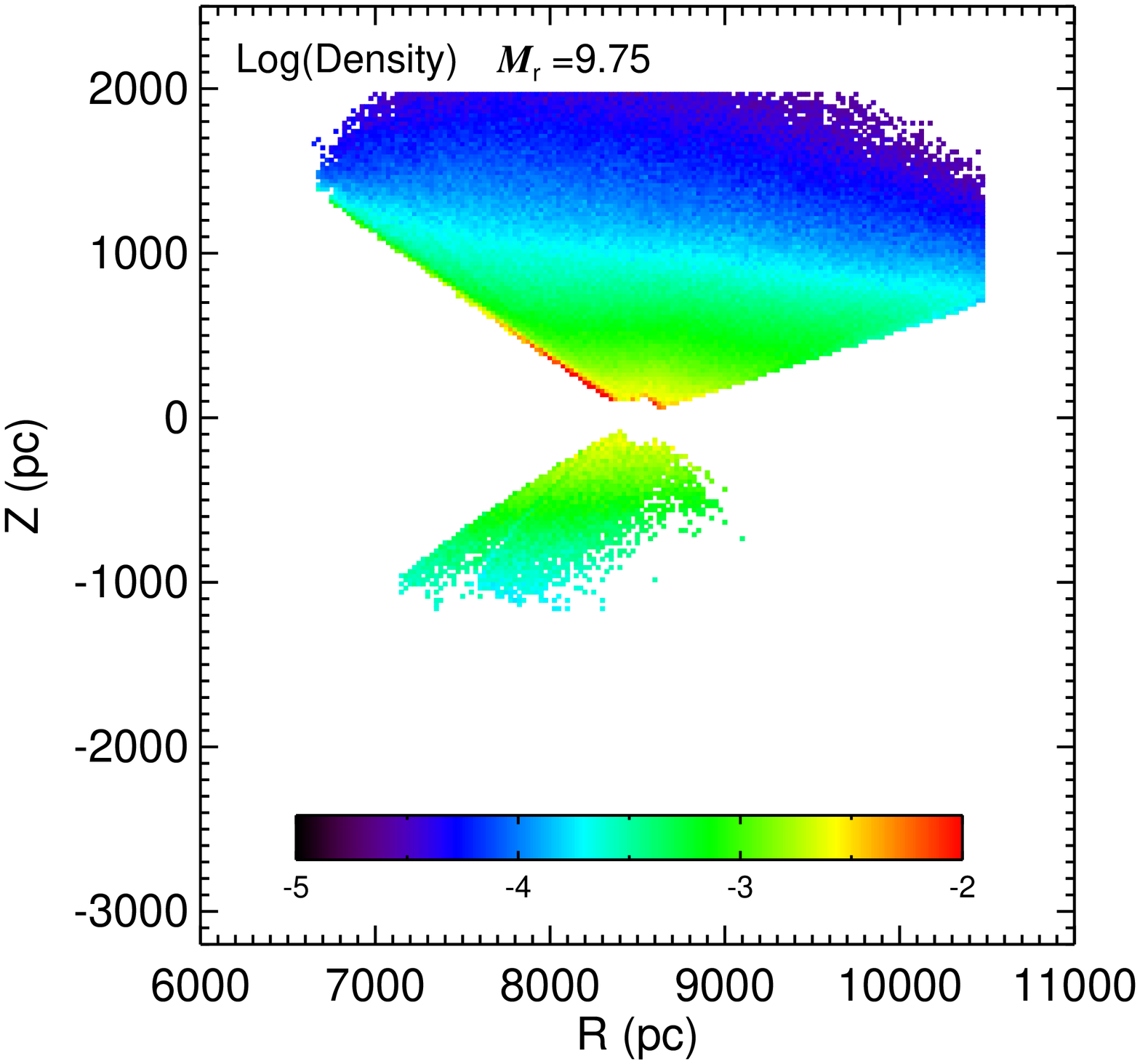}{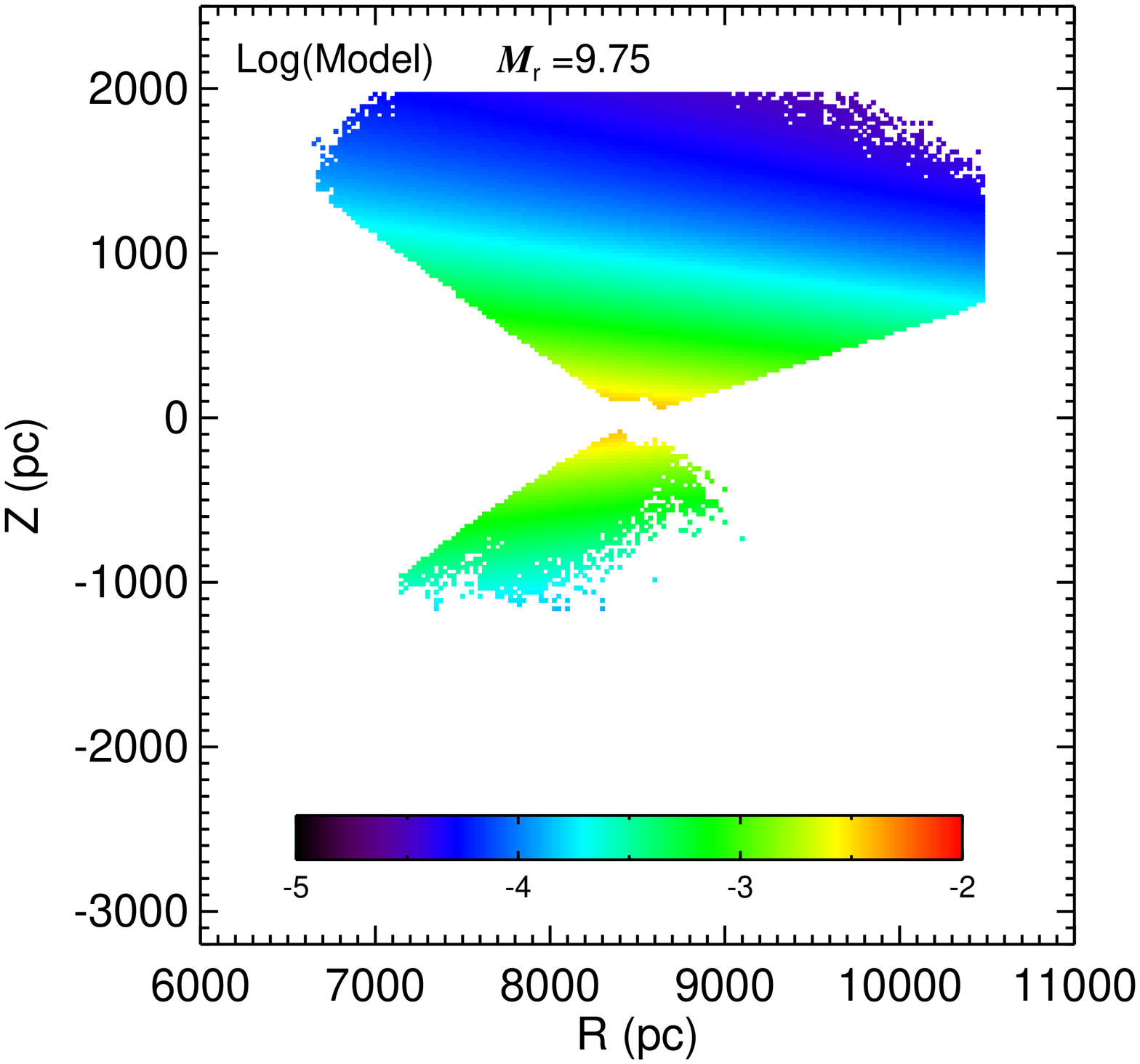}
 \caption{\textit{Left Panel} - The stellar density distribution in $R$
 and $Z$ for a 0.5 mag slice in $M_r$ centered on $M_r = 9.75$.  Over
 1 million stars are contained within this slice.  Note the smooth
 decrease in the vertical direction (due to the exponential scale
 heights of the Galactic thin and thick disks) and the signature of
 the Milky Way's scale length, manifested as a decrease in density
 with $R$.  \textit{Right Panel} - The corresponding two-disk model fit
 for the same $M_r$ slice.}
 \label{fig:maps}
\end{figure}

To address this issue, \cite{2010AJ....139.2679B} employed the following technique for
measuring the luminosity function.  First, absolute magnitudes were
assigned and distances to each star were computed using new $M_r, r-z$ and
$M_r, r-i$ color--absolute magnitude relations (CMRs) estimated from nearby
stars with $ugriz$ photometry and accurate trigonometric parallaxes.  Next, a small range in
absolute magnitude (0.5 mag) was selected and the stellar density was
measured as a function of Galactic radius
(\textit{R}) and Galactic height (\textit{Z}).  This range in
absolute magnitude was selected to provide high resolution of the LF,
with a large  number of stars ($\sim 10^6$) in each bin.  Finally, a
Galactic profile was fit to the $R,Z$ density maps, solving for the
shape of the thin and thick disks and the local stellar density.  The
LF was then constructed by combining the local density of each absolute magnitude slice.
An example of one density map and the corresponding fit is shown in
Figure \ref{fig:maps}.

\subsection{Systematic Challenges:  Binarity, Metallicity and Absolute
  Magnitudes}
The measured ``raw'' LF is subject to various
systematic effects.   Unresolved binarity, metallicity gradients
within the Milky Way,
interstellar extinction and Malmquist bias can all influence a star's
estimated absolute magnitude and distance.  Understanding these
effects is important, since the photometric data set contains
millions of stars and Poisson errors are no longer significant within a
given LF bin.  Rather, systematic errors in absolute magnitudes are
the dominant source of uncertainty.  In the \cite{2010AJ....139.2679B}
study, the uncertainty in each LF bin
was calculated by repeating the entire analysis multiple times, using
five different prescriptions for the CMR, which were selected to account for metallicity gradients,
extinction and reddening.

The effects of unresolved binarity and Malmquist bias were quantified
using Monte Carlo realizations of the Milky Way.
Each model was populated with synthetic systems that were consistent
with the observed Galactic structure and LF.  Binaries were included
within the simulations, using four different prescriptions for mass
ratio distributions.  The mock stellar catalog
was analyzed with the same pipeline as the actual observations.  The
differences between the input and ``observed'' Galactic structure and
LF were used to systematically correct the observed ``raw'' values
from SDSS data.  The single--star LF was also estimated during this step.
The detailed analysis of Malmquist bias and
unresolved binaries and their effect on the resultant LF are given in
\cite{2010AJ....139.2679B}.

\subsection{The SDSS Field LF and MF}
\begin{figure}[!ht]
 \plotone[width=0.5\textwidth]{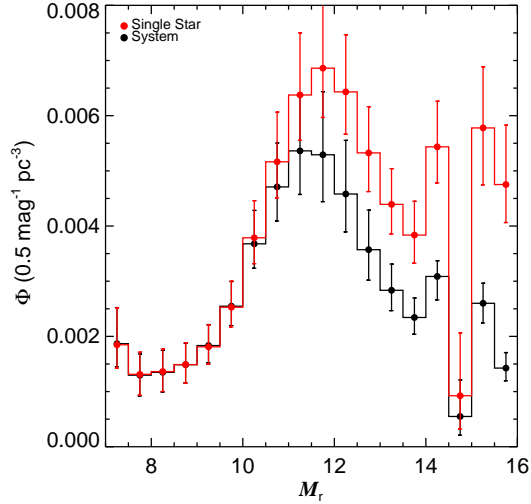}
 \caption{The single--star LF (red line) and system LF (black line)
   for SDSS M dwarfs.  The LFs have been corrected for the effects of
   unresolved binarity and Malmquist bias.}
 \label{fig:lf_sing_system}
\end{figure}

The final adopted system and single--star $M_r$ LFs are presented in
Figure \ref{fig:lf_sing_system}.  The uncertainty in each LF bin
was computed from the full spread due to differences among
CMRs and binary prescriptions.   The system LF (black
circles) rises
smoothly to $M_r \sim 11$, with a fall off to lower masses.  The
single--star LF (red circles) follows a similar track, but maintains a
density increase of $\sim 2\times$ compared to the system LF at the
same absolute magnitude.  This implies that lower luminosity stars are
easily hidden in binary systems, but isolated low--luminosity systems
are not common.  

\begin{figure}[!ht]
 \plottwo{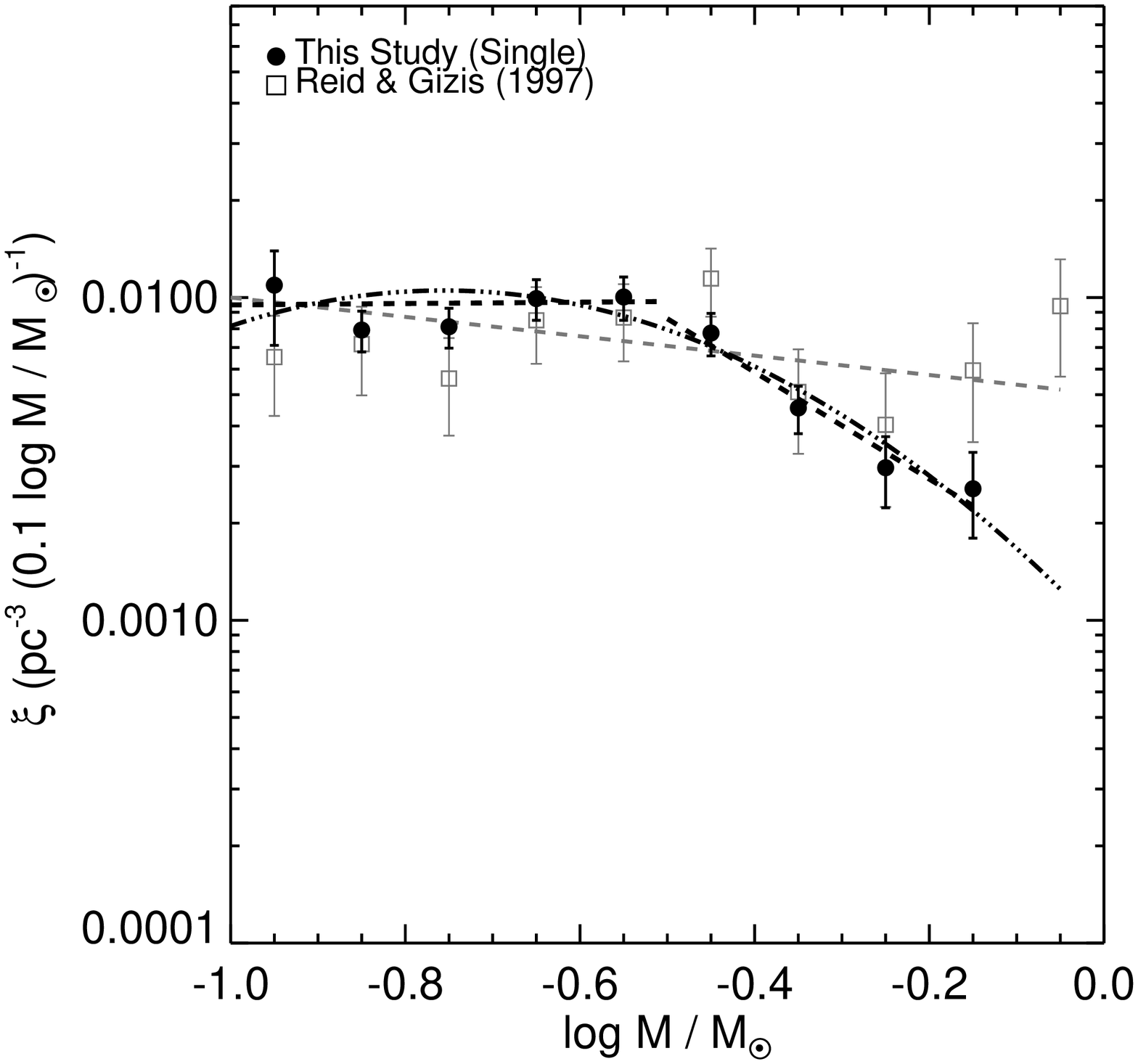}{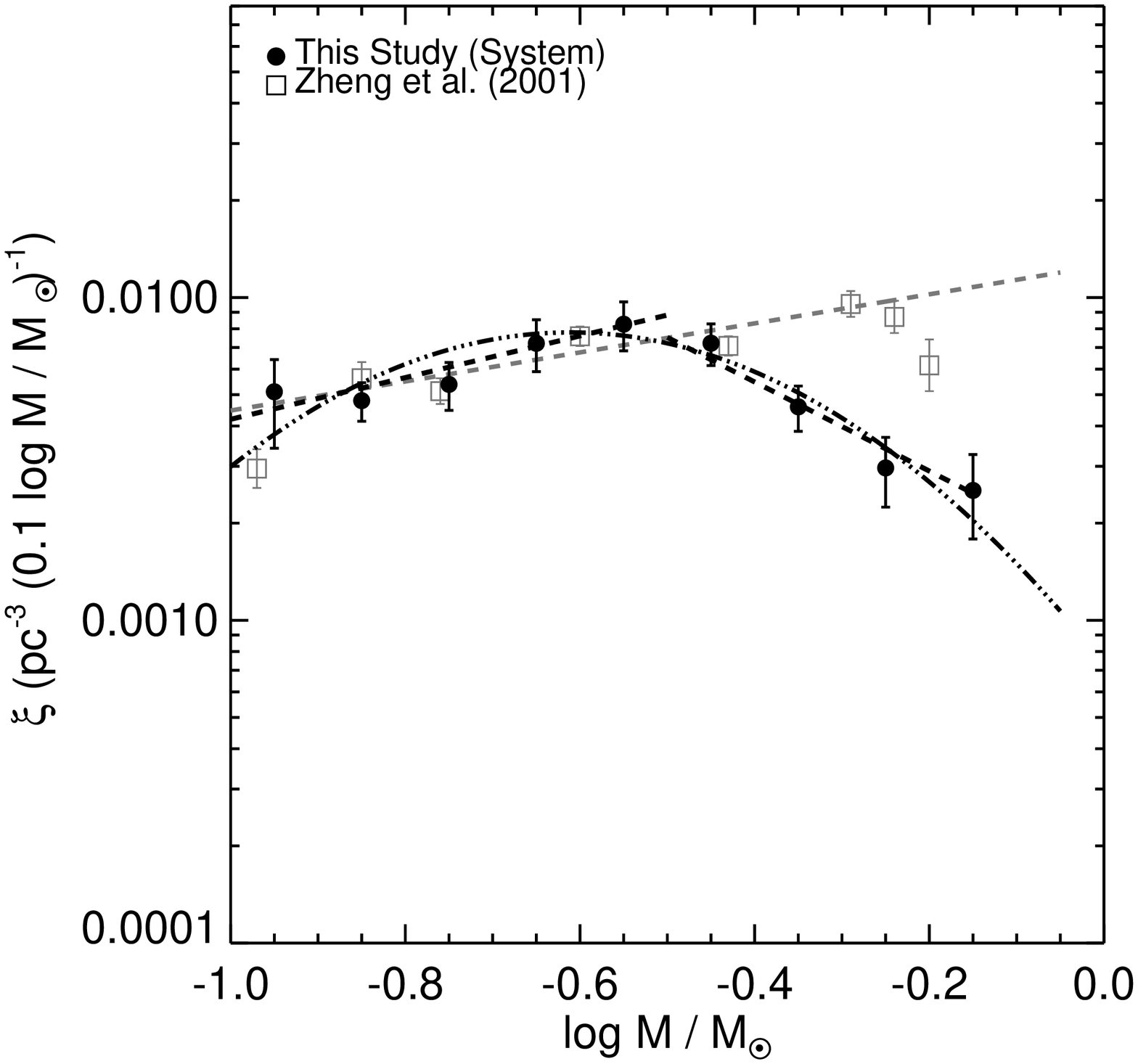}
 \caption{\textit{Left Panel} - The SDSS single--star MF (filled
     circles) compared to the local census of
     \cite{1997AJ....113.2246R}.  The two samples show agreement over
   a large range in mass, suggesting the correction for unresolved
   binarity is valid.  \textit{Right Panel}~- The SDSS system MF
   compared to the study of \cite{2001ApJ...555..393Z}.  The
   disagreement at large masses is most likely due to differences in
   the assumed CMRs.}
 \label{fig:reid_zheng}
\end{figure}

Using the mass--magnitude relations of \cite{2000A&A...364..217D}, the field
system and single--star LFs were converted to MFs, shown in Figure
\ref{fig:reid_zheng}.  The system MF exhibits similar behavior to the
LF, peaking near 0.3 $M_{\odot}$.  The single--star MF is compared to
the local ``eight parsec'' MF of \cite{1997AJ....113.2246R} in Figure
\ref{fig:reid_zheng}.  The \cite{1997AJ....113.2246R} sample consisted
of 558 stars spread over 3$\pi$ steradians.  There is broad
agreement over most of the mass range probed, with the SDSS MF falling
below the local MF at higher masses.  The discrepancy is probably
due to the assumed CMR, which may have overestimated the brightness
of higher mass M dwarfs \citep[see contribution by Hawley et al., this
volume; ][]{bochanski10}.  To first order, this would lead to
overestimated distances and underestimated local stellar densities.
Future investigations will rederive the LF and MF with updated
CMRs.  The SDSS MF was fit with two functions: a broken power
law and lognormal.  Both adequately represent the underlying data
points, but a lognormal is slightly preferred due to the smaller
number of free parameters.

The right panel of Figure \ref{fig:reid_zheng} compares the SDSS system MF with the MF
reported by \cite{2001ApJ...555..393Z}.  The \cite{2001ApJ...555..393Z} HST survey was a pencil
beam survey of 1,400 M dwarfs over $\sim$ 1 sq.\ deg..
Their system MF agrees with most of the SDSS MF, especially at lower
masses.  The \cite{2001ApJ...555..393Z} power law fit agrees favorably with the low--mass
component of the SDSS broken power law fit.  At higher masses, there is a
disagreement, which is likely due to CMR differences.

\begin{figure}[!ht]
 \plotone[width=0.5\textwidth]{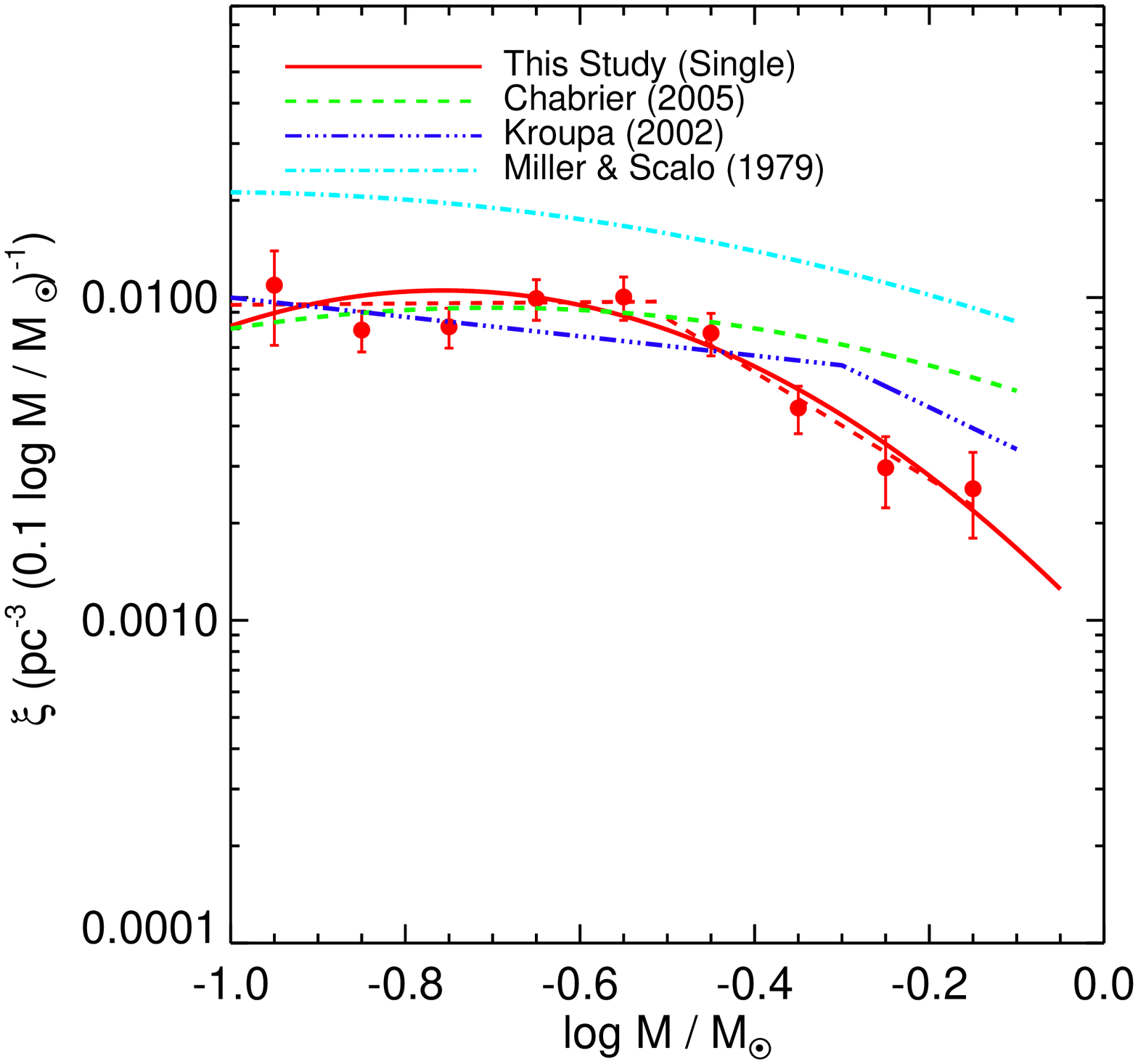}
 \caption{Shown is the MF data and best log-normal fit from
   \citet[solid circles and red line]{2010AJ....139.2679B}, along with the analytic MF fits of
   \citet[green dashed line]{2005ASSL..327...41C}, \citet[dark blue
   dash--dot--dot--dot line]{2002Sci...295...82K} and \citet[light
   blue dash-dot line]{1979ApJS...41..513M}.  I stress that comparing
   actual MF data is preferred to comparing analytic fits.}
 \label{fig:big_shots}
\end{figure}

When searching for variations between MFs, it is crucial that MF $data$ is
compared, rather than best--fit functional forms.  This point, first
introduced in \cite{covey08} and reiterated by \cite{2010ARA&A..48..339B} is
important.  The comparison of $fits$ and not $data$ may have
exaggerated the differences between the local
census \citep[i.e.][]{1997AJ....113.2246R} and distant samples \citep[i.e.][]{2001ApJ...555..393Z},
although other factors, such as unresolved binarity \citep[e.g.][]{1991MNRAS.251..293K,2003ApJ...586L.133C}
also contributed.  Finally, the SDSS single-star MF is compared to selected seminal works
in the field in Figure \ref{fig:big_shots}.  

\section{Kinematics}\label{sec:kinematics}
While photometric studies can be used to study the distribution of M
dwarfs throughout the Galaxy and infer their fundamental properties,
they offer a static glimpse of the population.  The dynamics of stars
through the Milky Way, as measured by radial velocities and proper
motions, can be used to infer the underlying local mass distribution and
gravitational potential.  In particular, kinematic studies of
M dwarfs have been useful for studying the properties of the
thin and thick disks.  
Using a sample of $\sim$ 7,000 M dwarfs along one line of sight,
\cite{2007AJ....134.2418B} examined the $UVW$ velocity distributions
as a function of distance from the plane.  Thin and thick disk
components were identified with the use of probability plots
\citep{1973PASP...85..573L, 1995AJ....110.1838R}.  In general,
both the thin and thick disks dispersions increase with height above
the Plane, with the $W$ velocity dispersion being the
smallest of the three directions.  Low--mass stars follow
the Galactic potential across the entire mass range.  In other words,
early and late--type M dwarfs, despite differing in mass by
nearly a factor of ten, have similar velocity dispersions at the
same height from the Galactic midplane.  Similar trends with height were also quantified by
\cite{2009AJ....137.4149F} using proper motions and distances of
$\sim$ 2 million SDSS M dwarfs.

\begin{figure}[!ht]
 \plotone[width=0.9\textwidth]{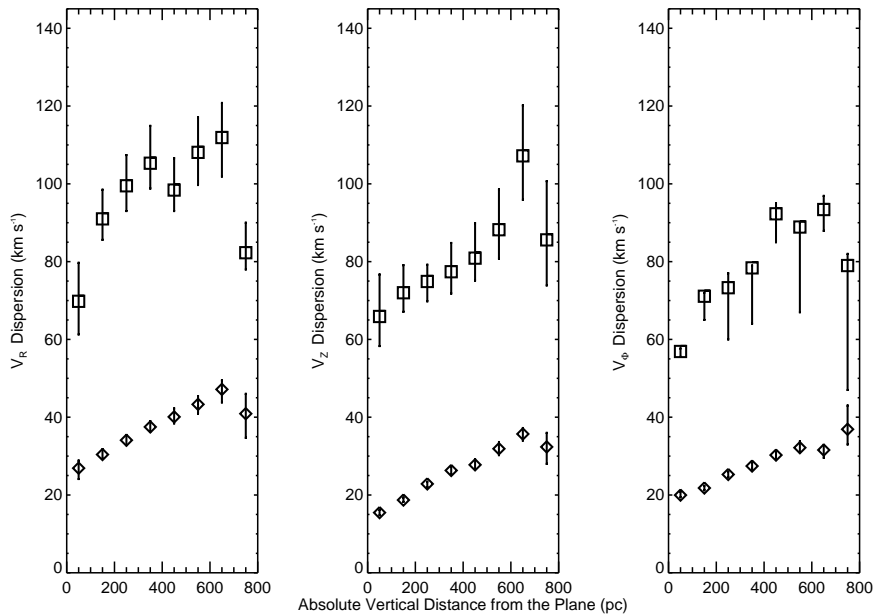}
 \caption{Thin disk (diamonds) and thick disk (open squares) velocity
   dispersions as a function of height in a Galactocentric coordinate system.  Note the
   general increase with height above the Galactic plane.  Figure from
 \cite{pineda}.}
 \label{fig:disps}
\end{figure}

The analysis from \cite{2007AJ....134.2418B} was expanded to the entire SDSS
footprint by \cite{pineda}, using the largest spectroscopic sample of
M dwarfs ever assembled \citep{west10}.  \cite{pineda} determined mean velocities,
dispersion and relative scalings between the thin and thick disk.
They searched for velocity gradients radially in
the disk, and observed a similar increase in dispersion away the
Plane  (see Figure \ref{fig:disps}).  \cite{pineda} kinematically estimated the local fraction of
thick disk stars to be $\sim 5 \%$, similar to the photometric result from 
\cite{2010AJ....139.2679B}.  This
study exemplifies the utility of low--mass stars as probes of the
local Milky Way.

\section{Low-Mass Stars in a Galactic Context}\label{sec:case_study}
In addition to their utility as tracers of the local Milky Way, low--mass dwarfs have
intrinsic properties, such as chromospheric activity
\citep{2004AJ....128..426W,2007AJ....134.2418B,2007AJ....133.2258S},
that can be examined in a broader context.  Additionally, metallicity
may be studied using subdwarfs \citep{2003AJ....125.1598L}, readily
identified by their spectra, which show enhanced calcium hydride (CaH)
absorption.  Subdwarfs have been easily detected in large-scale
surveys such as the SDSS \citep[e.g., ][]{2004AJ....128..426W,2008ApJ...681L..33L}. 

The study of M dwarf chromospheric activity in a
Galactic context is described below.  Similar investigations were severely limited before the
advent of large spectroscopic databases, such as SDSS.  It highlights
the importance of studying M dwarfs in a broader context within the
Milky Way, as their distributions within the Galaxy may be used as a
valuable constraint on intrinsic properties. 

\subsection{Magnetic Activity: A Case Study}
Low--mass stars harbor strong magnetic dynamos
\citep{2008ApJ...676.1262B}, which in turn heat their chromospheres,
producing Balmer line emission.  For early-type stars (M0-M4), the
dynamos are probably powered by stellar rotation, but the driving mechanism at
later types is not well known.  Ca H $+$ K emission is also a
tracer of activity in M dwarfs \citep{2009AJ....137.3297W}, but is often not observationally
accessible for these dim, red stars.  Prior to SDSS, surveys that
measured the fraction of active M dwarfs often consisted of $<$ 1000 stars,
with very few stars ($< 100$) in the faintest luminosity bins
\citep{1996AJ....112.2799H, 2000AJ....120.1085G}.  Since previous samples were flux--limited, they were heavily
biased towards observing nearby, young stars.

The first major study of activity in M dwarfs with SDSS data was performed by
\cite{2004AJ....128..426W}.  This study observed many stars at distances 5-10 times larger than
previous studies due to the multi-fiber capabilities of the SDSS
spectrographs.  The differences between the \cite{2004AJ....128..426W} study and
the past surveys were immediately obvious.  First, some spectral
types (M7 and M8) which were previously observed to have active fractions near
100$\%$ had lower active fractions, near 70$\%$.  The active fraction
distributions for all spectral types declined with distance from the
Galactic midplane with a spectral type dependent slope.
\cite{2004AJ....128..426W} suggested this was an age effect, as stars further from
the Plane also exhibited larger velocity dispersions.  Thus, older
stars have undergone more dynamic interactions and also possess
weaker magnetic fields.  The M dwarf age-activity relation was
calibrated by \cite{2008AJ....135..785W}, which determined the activity lifetime
for M dwarfs as a function of spectral type.  By using a simple model
that fit both the velocity and active
fraction distributions within the Galaxy, \cite{2008AJ....135..785W}
were able to constrain the ages of stars, a 
fundamental stellar property.  It is worth noting that without large
samples of M dwarfs, such as those available with SDSS, this type of
analysis would not be possible.  The \cite{2004AJ....128..426W} study also highlights the
potential biases that are encountered when limiting samples to the solar neighborhood.  The Sun is very close to the
midplane \citep[15 pc; ][]{1995ApJ...444..874C}, and nearby stars are
predominately young \citep[1-2 Gyrs;][]{2004A&A...418..989N}, biasing local samples.

\subsection{Metallicity \& Stellar Ages:  The Next Frontiers}
Current models of low--mass stars cannot accurately reproduce
observable features.  This has led to uncertainty in determining
fundamental parameters, such as mass, age and luminosity.  Thus,
empirical relations have been developed to estimate these properties.
Mass--magnitude relations determined from eclipsing binaries
\citep{2000A&A...364..217D}, have proven extremely useful, but other fundamental
parameters, most notably age and metallicity have not been fully
calibrated.

Recent advances in measuring metallicity are promising \citep[e.g.][and see the review of the 
metallicity splinter
session in this volume]{2007ApJ...669.1235L,2009ApJ...699..933J,2010ApJ...720L.113R}.  In particular, \cite{2010ApJ...720L.113R} have calibrated an infrared technique that
delivers 0.15 dex precision from $K$ band spectra.  However, for the
vast majority of M dwarfs with optical spectra, a similar level of precision has
not been achieved.  Further observations of M dwarfs with known
metallicities (i.e., those in wide binaries with FGK companions \citep[i.e.,][]{2010AJ....139.2566D} or
cluster members) need to be obtained before a suitable calibration can
be determined.  However, once precise metallicities can be measured for
M dwarfs, they will be a powerful probe of the Milky Way's chemical
history.

Age is notoriously difficult to estimate for field stars (of any
mass).  However, age--activity relations \citep[i.e.][]{2008AJ....135..785W},
suggest that observational tracers, such as chromospheric activity,
can constrain the ages of M dwarfs to a few Gyr.  Further
work is needed to produce precise estimates of M dwarf
ages.  The next generation of synoptic surveys, such as LSST and
PanSTARRS may be useful in this endeavor and provide a link between
rotation periods and stellar age.

\section{Conclusions}\label{sec:conclusions}
I have briefly detailed some of the investigations that examine both
intrinsic M dwarf properties and Galactic structure and
kinematics.  The requisite data for such studies, namely
precise, deep, multi-band photometry over large areas of the sky, will
become more common in the future, as additional large surveys come online.
Within the next decade, surveys such as PanSTARRS, LSST and GAIA
\citep{2002SPIE.4836..154K, 2008arXiv0805.2366I, 2001A&A...369..339P},
will produce new data sets that are ideal for similar investigations.

While databases containing millions of low--mass stars will be the
norm for future studies, it is important to note that systematic
errors will be the dominant source of error in many areas.  In
SDSS, nearly all the observed M dwarfs do not have reliable trigonometric
parallaxes, forcing the use of photometric parallax estimates.
Reliably calibrating and testing these photometric parallaxes will
only become more important as surveys push to larger distances.  Other
systematics, such as how metallicity affects the absolute magnitude of
M dwarfs, will also be crucial to future
investigations.
 
Finally, the importance of SDSS spectroscopy can not be overstated.
The large spectroscopic samples of SDSS M dwarfs have
enabled many novel investigations.  Significant
spectroscopic followup of the next generation of surveys should
be a high priority.

\acknowledgements I would like to thank the Cool Stars 16 SOC for the
opportunity to present this work.  I also thank Suzanne Hawley, Andrew
West and Kevin Covey for their insight and many enlightening
conversations over the years.  I thank Sebastian Pineda for kindly
providing a figure.  I would also like to acknowledge the
comments of Neill Reid, Ivan King, Niall Deacon and Gilles Chabrier, who all
contributed to work presented in this article.


\end{document}